\documentclass{phyeauth}
\usepackage{graphicx}
\usepackage{amsmath}
\usepackage{amssymb}

\def\prl{Phys.\ Rev.\ Lett.\ }
\def\prb{Phys.\ Rev.\ B\ }
\def\rmp{Rev.\ Mod.\ Phys.\ }

\begin{document}

\begin{frontmatter}

\title{
Dynamics of disordered quantum Hall crystals
}

\author{Michael M. Fogler} 

\address{
University of California San Diego,
Department of Physics,
9500 Gilman Drive, La Jolla, CA 92093-0319, USA}

\begin{abstract}

Charge density waves are thought to be common in two-dimensional
electron systems in quantizing magnetic fields. Such phases are formed
by the quasiparticles of the topmost occupied Landau level when it is
partially filled. One class of charge density wave phases can be
described as electron solids. In weak magnetic fields (at high Landau
levels) solids with many particles per unit cell - bubble phases -
predominate. In strong magnetic fields (at the lowest Landau level) only
crystals with one particle per unit cell - Wigner crystals - can form.
Experimental identification of these phases is facilitated by the fact
that even a weak disorder influences their dc and ac magnetotransport in
a very specific way. In the ac domain, a range of frequencies appears
where the electromagnetic response is dominated by magnetophonon
collective modes. The effect of disorder is to localize the collective
modes and to create an inhomogeneously broadened absorption line, the
pinning mode. In recent microwave experiments pinning modes have been
discovered both at the lowest and at high Landau levels. We present the
theory of the pinning mode for a classical two-dimensional electron
crystal collectively pinned by weak impurities. We show that long-range
Coulomb interaction causes a dramatic line narrowing, in qualitative
agreement with the experiments.

\end{abstract}

\end{frontmatter}

%
\section{Introduction}
\label{Sec:Introduction}

Two-dimensional electron gas (2DEG) in quantizing magnetic fields is
known to be a system with a rich phase structure. Recently, much
attention has been devoted to charge-density wave phases, which include
stripe phase, bubble phase, and a Wigner
crystal~\cite{Das_Sarma_book,Eisenstein_01,Fogler_review}. An important
new information on the dynamics of such phases was provided by a group
of microwave experiments~\cite{Li_97,Beya,Li_00,Ye_02,Lewis_02}. They
demonstrated that the low-frequency ac response of
magnetic-field-induced charge-density waves is strongly affected by
disorder, in ways not anticipated in prior theoretical work on the
subject. Motivated by these intriguing results, Huse and the present
author~\cite{Fogler_00} reconsidered the problem of collective dynamics
of a Wigner crystal pinned by quenched disorder. Here I give a brief
account of this work and its more recent extensions. Much of the
foregoing discussion also applies to the bubble phase. The rest of this
section is devoted to a brief introduction to the charge-density wave
phases, conventional expectations regarding the response of a pinned
system, and how they compare with the observed behavior.
Sections~\ref{Sec:Model} and \ref{Sec:Analysis} outline our theoretical
model and its analysis. Finally, Sec.~\ref{Sec:Comparison} contains
comparison of our theory with the experiment and concluding remarks.


The bubble phase~\cite{Fogler_review} can be viewed as a
crystal made of multi-electron droplets (bubbles). The number of
particles per bubble $M$ depends on the Landau level filling fraction
$\nu$ and changes in discrete steps at certain critical values of $\nu$.
Here $M$ counts only the quasiparticles of the topmost ($N$th) Landau
level, which is partially filled. The other Landau levels are completely
filled and in a first approximation, provide an inert background of
constant uniform density. Bubble phases with $M > 1$ appear when three
or more Landau levels are populated, $\nu > 4$ ($N \geq 2$).
The Wigner crystal is the $M = 1$ bubble phase (one particle per unit cell).
It forms at any Landau level $N$ provided $\nu$ is close enough
to an integer.
We refer to the bubble and the Wigner crystal phases jointly as
electron crystals. At zero temperature, in the absence of disorder and
commensurability effects such crystals have a perfect triangular lattice
and a long-range order. They possess collective modes referred to as
magnetophonons. Wigner crystal has a single magnetophonon
branch~\cite{Bonsall_77}, bubble phases may have
several~\cite{Fertig_xxx}; however, in both cases only one mode is
gapless. The following discussion is devoted exclusively to the dynamics
of the gapless mode. Quasiclassically, a magnetophonon excitation can be
understood as a coherent precession of $M$-electron bubbles around their
equilibrium positions, with the charge distribution (form-factor) of
individual bubbles being fixed. In this picture the dynamical degrees of
freedom are the centers of mass of the bubbles.

\begin{figure}[h]
\begin{center}
\includegraphics[width=2.0in,bb=101 431 477 679]{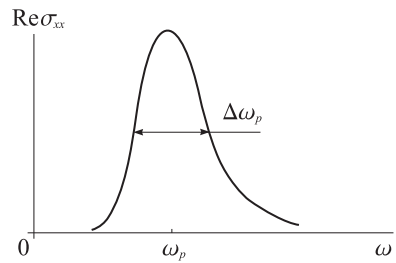}
\caption{
The sketch of the absorption line due to pinning.
\label{Fig_pinning_mode}
}
\end{center}
\end{figure}

The effects of disorder on gapless magnetophonon excitations belong to a
broad class of phenomena known as {\it pinning\/}. The pinning affects a
variety of physical systems. It has been extensively studied in the
context of conventional charge-density waves~\cite{Gruner_88} and vortex
lattices in superconductors~\cite{Blatter_94}. Two pinning mechanisms
are distinguished, the individual and the collective ones. In this paper
we focus on the latter. The collective pinning is the regime where the
random potential due to impurities and other defects is too weak to
significantly deform the crystalline lattice, so that the crystal is
well ordered at small length scales. However, the cumulative effect of
the disorder eventually dominates the crystal elasticity at a pinning
length, which is much larger than the lattice constant $a$. The
long-range order of the crystal is thereby destroyed. Similarly, the
low-frequency collective modes are unaffected by disorder at the scale
of $a$, but get localized at some larger scale. These localized modes
give rise to a maximum at a disorder-dependent frequency $\omega_p$ in
the real part of diagonal conductivity $\Re{\rm e}\sigma(\omega)$. Such
maxima (pinning modes) have been studied primarily in conventional
charge-density wave materials~\cite{Gruner_88,CDW_pinning_mode}.
However, a few observations of similar absorption lines due to 2DEG
subject to a strong magnetic field have also been
reported~\cite{Palaanen_92}. The aforementioned latest experiments on
high-mobility 2DEG provided much more extensive and higher resolution
data in the strong-field regime~\cite{Li_97,Beya,Li_00,Ye_02} and also
detected pinning resonances in weak fields~\cite{Lewis_02}, which were
interpreted as signatures of bubble phases and Wigner crystals at higher
Landau levels. The quantities that can be extracted and analyzed most
reliably are (i) the position $\omega_p$ of the peak in $\Re{\rm
e}\,\sigma$, (ii) its width $\Delta\omega_p$, and (iii) the area $F$ under
the curve (see Fig.~\ref{Fig_pinning_mode}).

\begin{figure}[h]
\begin{center}
\includegraphics[width=2in,bb=190 428 390 546]{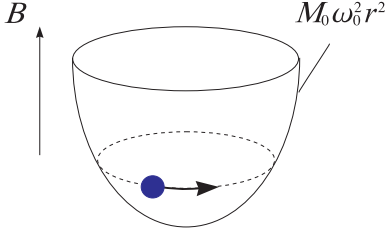}
\caption{
Naive cartoon of a localized magnetophonon mode. The ball represents a
single domain of the pinned crystal, which has mass $M_0$. The
paraboloid symbolizes the pinning potential in which this domain
resides. The dashed line shows the drift trajectory in the
two-dimensional plane ${\bf r} = (x, y)$. 
\label{Fig_bowl}
}
\end{center}
\end{figure}

(i) The dependence of $\omega_p$ on magnetic field and disorder is often
discussed in the framework of a popular physical
picture~\cite{Fukuyama_78,Normand_92} illustrated by Fig.~\ref{Fig_bowl}. Each
localized magnetophonon mode is visualized as precession (or drift) of a
crystallite with linear dimension of the order of the pinning length
$R_c$ in a potential well created by disorder. Let $\omega_c = e B / m_e
c$ be the cyclotron frequency in the external magnetic field $B$. For a
potential well with curvature $\sim M_0 \omega_0^2$, where $M_0$ is the
total mass of the patch, an elementary calculation gives the frequency
of the drift motion $\omega_p \sim \omega_0^2 / \omega_c$ in the
experimentally relevant case $\omega_c \gg \omega_p$. If $\omega_0$
depends only on the amount of disorder (e.g., impurity concentration)
but not on $B$, then $\omega_p \propto 1 / B$.

(ii) As far as $\Delta\omega_p$ is concerned, in a random system of
independent oscillators we expect a broad distribution of
$\omega_0^2$~\cite{Normand_92}, which implies a broad spectrum of their
eigenfrequencies, i.e., $\Delta\omega_p \sim \omega_p$.

(iii) Finally, for the integrated oscillator strength $F$, this model
predicts~\cite{Fukuyama_78}, $F = (\pi / 2) (e^2 n_e / m_e) (\omega_p /
\omega_c)$, where $n_e$ and $m_e$ are the electron concentration and
effective mass, respectively. 

Despite a certain appeal of these arguments, their predictions are not
supported by experiments. First, the pinning frequency shows a
complicated nonmonotonic dependence on $B$. Second, the absorption line
is very narrow. Quality factors $Q = \omega_p / \Delta \omega_p$ as
large as ten were reported in the published literature on the high-field
Wigner crystal~\cite{Li_97,Li_00,Ye_02} and lower but still respectable
quality factors $Q \sim 3$ for the bubbles. Only the integrated
oscillator strength $F$ is in a rough agreement~\cite{Lewis_02}
with the simple theory.

It turns out that the anomalous $B$-dependence can in principle be
explained by the interplay between the correlation length of the pinning
potential and the $B$-dependent form-factor of the electron
wavefunctions~\cite{Chitra_98,Fertig_99,Fogler_00}. On the other hand,
to explain the small linewidth one has to develop a theory that goes
beyond the cartoon depicted in Fig.~\ref{Fig_bowl}. Such a theory is the
subject of this paper. We will show that the line narrowing
is due to the long-range Coulomb interaction in the electron crystal.

\section{Definition of the model and the pinning frequency}
\label{Sec:Model}

We treat Wigner and bubble phases on the same footing, as crystals of
{\it classical\/} particles of mass $M m_e$, charge $M e$, and average
concentration, $n_e / M$. They interact with each other via potential
$V({\bf r})$ and are subject to a weak random potential $U({\bf r})$. We
assume that $V$ and $U$ incorporate the appropriate form-factors,
calculated using the quantum wavefunctions of the
bubbles~\cite{Das_Sarma_book,Fogler_review}. (This is the only place
where quantum mechanics enters our model.) We describe the deviation of
the crystal lattice from the ideal periodicity in terms of the elastic
displacement field ${\bf u}_{e l} = {\bf u}^{(0)} + {\bf u}$, which we
split into the static ground-state distortion ${\bf u}^{(0)}({\bf r})$
and the magnetophonon fluctuations ${\bf u}({\bf r}, t)$. We restrict
ourselves to the harmonic approximation, where magnetophonon modes are
obtained by diagonalizing the dynamical matrix
%
$
[{\bf D}]^{-1} = [{\bf D}^0]^{-1} \delta_{{\bf q}_1, {\bf q}_2}
+ \delta\tilde{\bf S}({\bf q}_1 - {\bf q}_2),
$
%
where ${\bf D}^0 = {\bf D}^0({\bf q}, \omega)$ is a
dynamical matrix of a uniformly pinned crystal. In the convenient basis
of transverse (T) and longitudinal (L) components, ${\bf u}({\bf q}) =
\hat{\bf q} u_L({\bf q}) + [\hat{\bf z} \times \hat{\bf q}] u_T({\bf
q})$, where $\hat{\bf q} = {\bf q} / |{\bf q}|$, ${\bf D}^0$ has the
form~\cite{Fukuyama_78,Bonsall_77}
\begin{equation}
     [{\bf D}^0]^{-1} = \left[
\begin{array}{cc}
\mu q^2 + S_0 - \rho \omega^2 & -i \rho \omega \omega_c\\
i \rho \omega \omega_c        & \lambda q^2 + S_0 - \rho \omega^2
\end{array}\right],
\label{D_0}
\end{equation}
where $\rho = m_e n_e$ is the average mass density, $\mu\sim M^2 e^2 /
\kappa a^3$ is the shear elastic modulus, $\lambda = \lambda(q) = 2 \pi
e^2 n_e^2 / \kappa q$ is the effective bulk modulus, and $\kappa$ is the
dielectic constant of the medium. The $q$-dispersion of $\lambda$ is due
to long-range Coulomb interaction. At small $q$, we have $\lambda(q) \gg
\mu$, i.e., compressions cost much more energy than shear deformations.
Parameter $S_0$ is the average value of the diagonal components of the
curvature tensor ${\bf S} = \nabla\nabla U({\bf r})$ evaluated at the
ground-state particle positions. From standard collective pinning
arguments, we expect $S_0 \sim \mu / R_c^2 \equiv \rho\omega_0^2$, where
$R_c$ is the pinning length related to the amplitude of the random
potential $U$ and the shear modulus $\mu$~\cite{Fogler_00}. The quantity
$\delta\tilde{\bf S} \equiv \tilde{\bf S} - S_0 {\bf I}$ describes
fluctuations of the curvature, where ${\bf I}$ is the identity matrix
and tilde denotes the Fourier transform. Our goal is to calculate the ac
conductivity, ${\sigma}(\omega) = -i e^2 n_e^2 \omega \langle{\bf
D}(\omega + i 0)\rangle$, where $\langle\ldots\rangle$ denotes disorder
averaging. The conductivity can be expressed in terms of the self-energy
%
$
{\bf \Pi} ({\bf q}, \omega) \equiv S_0 {\bf I} +
\left(\langle{\bf D}\rangle^{-1} - [{\bf D}^0]^{-1}\right),
$
%
as follows:
\begin{equation}
\Re{\rm e}\,\sigma(\omega) = e^2 n_e^2 \omega\, \Im{\rm m}\,
\frac {\Pi(0, \epsilon) - \epsilon}
{[\Pi(0, \epsilon) - \epsilon]^2 - \epsilon \epsilon_c}.
\label{sigma_xx}
\end{equation}
Here we introduced convenient ``energy'' variables
%
$
 \epsilon \equiv \rho \omega^2
$ and $
 \epsilon_c \equiv \rho \omega_c^2.
$
%
From Eq.~(\ref{sigma_xx}) one can see that in strong magnetic fields the
conductivity as a function of $\omega$ has a maximum of width
$\Delta\omega_p = -{\Im{\rm m}\,\Pi(0, \rho\omega_p^2)}/{\rho \omega_c}$
centered at $\omega_p = {\Re{\rm e}\, \Pi}/{\rho \omega_c}$. Up to
logarithmic factors~\cite{Fogler_00}, ${\Re{\rm e}\, \Pi} \sim S_0$, and
so $\omega_p = \omega_0^2 / \omega_c$. Thus, the position of the
conductivity peak is simply related to the pinning length $R_c$, in
basic agreement with the arguments given in Sec.~\ref{Sec:Introduction}
(and as mentioned above, $R_c$ can be straighforwardly expressed in
terms of the microscopic parameters of the model). The lineshape and
linewidth require much more elaborate analysis given in the next
section.

\section{Lineshape and linewidth of the pinning mode}
\label{Sec:Analysis}

To find $\Pi$ and $\sigma$ we have to resort to certain approximations,
which we motivate by physical arguments. As a starting point, consider a
simpler model of a pinned one-dimensional (1D) electron crystal with
short-range (screened) interaction $V$ in the absence of any magnetic
field. It has been known from early work~\cite{Feigelman,Giamarchi_88}
that $\Re{\rm e}\,\sigma(\omega)$ has a broad maximum at the
characteristic pinning frequency $\omega_p^{1D} = v / R_c$, where $R_c$
is the pinning length and $v$ is the sound velocity. However, there has
been a disagreement about the behavior of $\sigma$ at small $\omega$.
Recently, the present author~\cite{Fogler_02} and later also Gurarie and
Chalker~\cite{Gurarie_02} advanced analytical and numerical arguments
that $\Re{\rm e}\,\sigma(\omega) \propto \omega^4$ for $\omega \ll
\omega_p^{1D}$. The physical picture is as follows~\cite{Fogler_02}. In
1D, elastic displacement $u$ is a scalar and there is a formal analogy
between the phonon problem and 1D localization in a random potential
$S(x) = U^{\prime\prime}(x)$. Indeed, the phonon mode with ``energy''
$\epsilon_i$ must satisfy the Schr\"odinger equation $-\rho v^2
u_i^{\prime\prime} + S(x) u_i(x) = \epsilon_i u$. Borrowing standard
arguments from the 1D localization literature, one can show that phonon
eigenmodes with $\omega \lesssim \omega_p^{1D}$ have localization length
$\sim R_c$. Each mode can be viewed as harmonic oscillations of a
segment of length $R_c$ in a potential well created by the pinning
centers. This echoes the picture depicted in Fig.~\ref{Fig_bowl} and
arguments given in Sec.~\ref{Sec:Introduction}. The $\omega^4$-law for
the the low-frequency (``soft'') modes follows from a careful derivation
of the the statistical distribution of the curvatures of such collective
potential wells. One finds~\cite{Fogler_02} that the density of states
$\nu = \langle \delta(\epsilon - \epsilon_i)\rangle$ of the eigenmodes goes
as $\nu(\epsilon) \propto \epsilon^s$ with $s = 3/2$. The conductivity
is proportional to $\nu$, and once the Jacobian of the transformation
from $\epsilon$ to $\omega$ is taken into account, one obtains $\Re{\rm
e}\,\sigma \propto \omega^{2 s + 1} = \omega^4$. What is important here
is that the derivation of $\nu$ and the exponent $s$ is rather
insensitive to microscopic details. Therefore, $\omega^4$-law is
expected to hold in higher dimensions and for other types of pinning
models~\cite{Aleiner_93,Gurarie_02}. Shortly below, we will make use of
this universality.

Let us now return to the 2D case but first examine a model, which is a
truncated version of the original one. Consider what happens if we
remove by hand all elements of the dynamical matrix ${\bf D}$ that
connect L and T modes. (On some crude level, it corresponds to a 2D
crystal in zero magnetic field.) As in 1D case, we reduce the problem to
solving a Schr\"odinger equation. Now we have
two such equations,
\begin{equation}
{\bf H}_T u_T \equiv -\mu \nabla^2 u_T + S_T({\bf r}) u_T = \epsilon u_T
\label{H_T}
\end{equation}
for T-modes and $-\lambda \nabla^2 u_L + S_L({\bf r}) u_L = \epsilon
u_L$ for L-modes. Here $S_T$ and $S_L$ are the T-T and L-L components of
the matrix ${\bf S}$, and $\lambda$ should be inderstood as integral
operator. What is the structure of the resultant eigenmodes? The T-modes
with $\omega < \omega_0$ again form a set of oscillators localized on the
scale of the pinning length $R_c$ ($\omega_0$, introduced in
Sec.~\ref{Sec:Model}, can be thought of as the $B = 0$ pinning
frequency). The soft T-modes correspond to $\omega \ll \omega_0$, i.e.,
$\epsilon \ll \rho\omega_0^2 \equiv \Pi_0$. Turning to the L-modes, in
principle, we also obtain a set of local oscillators. If $\lambda$ could
be replaced by some $q$-independent constant much larger than $\mu$, as
in the case of Coulomb interaction screened by a nearby metallic gate,
then the localization length of L-modes would be
$R_L = R_c \sqrt{\lambda / \mu}$.
It is much larger than $R_c$ because $S_L$ has about the same rms
fluctuations as $S_T$ yet the L-modes are much stiffer and resist
localization effects more efficiently. If $q$-dispersion of $\lambda$ is
taken into account, then one finds~\cite{Fogler_00} that the eventual
localization does occur, but at a scale so large that it can be treated
as infinite for all practical purposes. In this approximation,
randomness in $S_L$ does not affect the dynamics, so that $S_L$ can be
replaced by $S_0$. We arrived at a system of localized T-phonons and
delocalized L-phonons. In such a system, $\Re{\rm e}\,\sigma$ has an
infinitely sharp pinning mode. 

Let us now reinstate the L-T mixing. We will show that localization
does occur at some reasonable scale $R_M$ in this case and that
$\Delta\omega_p$ is finite albeit much smaller than $\omega_p$. However,
this is not a trivial task because the dynamical equations become much
more complicated. Fortunately, in the large-$B$ limit there are
two simplifications. First, we only need to consider mixing due to the
Lorentz force. (The L-T mixing due to disorder is not important for the
main results.) Second, the inertial terms $-\rho\omega^2$ in ${\bf D}^0$
[Eq.~(\ref{D_0})] can be neglected. The equation for $u_L$ becomes
\begin{equation}
(-\lambda \nabla^2 + S_0) u_L
 = \epsilon \epsilon_c [{\bf H}_T]^{-1} u_L.
\label{u_L_equation_I}
\end{equation}
The resolvent of the operator ${\bf H}_T$ [Eq.~(\ref{H_T})] on the
right-hand side of Eq.~(\ref{u_L_equation_I}) can be written in terms of
localized T eigenmodes discussed above. We will see that $R_M \gg R_c$;
therefore, it is legitimate to coarse-grain the model on the scale
$R_c$, so that each localized oscillator becomes represented by a single
site. This leads to the equation
\begin{equation}
(-\lambda \nabla^2 + \epsilon_0) u_L
 = \sum_j ({\epsilon \epsilon_c}/{\epsilon_j}) R_c^2
                 \delta({\bf r} - {\bf R}_j) u_L.
\label{u_L_equation_II}
\end{equation}
In the real space, it reads
\begin{equation}
u_L({\bf r}) = {\epsilon \epsilon_c} \sum_j
\frac{u_L({\bf R}_j)}{\epsilon_j} K({\bf r} - {\bf R}_j),
\label{u_L_equation_III}
\end{equation}
where $K(r)$ is the inverse Fourier transform of $R_c^2 / [\lambda(q)
q^2 + \epsilon_0]$. Function $K$ is of the order of $K_0 = R_c^2 / \lambda$
at $r \sim R_c$. At large $r$, it decays as $1 / r^3$.

The spectrum of eigenmodes can be obtained from
Eq.~(\ref{u_L_equation_III}) by setting ${\bf r} = {\bf R}_j$ and
solving the resultant system of linear equations. For example, if we
neglect randomness and replace $\epsilon_j$'s by their typical value
$\Pi_0$, we obtain $\epsilon \sim \Pi_0 / \epsilon_c$, and so $\omega_p
\sim \omega_0^2 / \omega_c$ in agreement with Sec.~\ref{Sec:Model}.

\begin{figure}[h]
\begin{center}
\includegraphics[width=2in,bb=108 461 318 651]{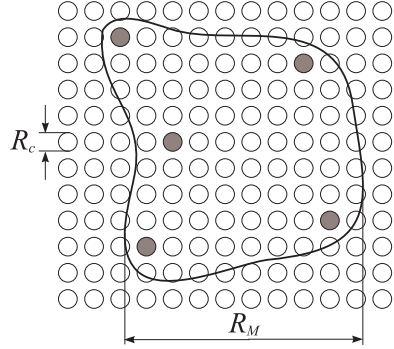}
\caption{
Schematic structure of
the localized magnetophonon eigenstate. Circles represent the T-modes.
Soft modes that act as strong scatterers are shaded.
\label{Fig_mode}
}
\end{center}
\end{figure}

Let us now try to account for the randomness. Consider a mode with
$\omega \sim \omega_p$. It is easy to see that soft modes with energies
$\epsilon_j \ll \epsilon_s = (\mu / \lambda) \Pi_0$ have a dramatic
effect on $u_L$. Within a distance $\sim (\lambda / \mu) R_c$ to the
locations of such a mode, $u_L$ is strongly suppressed. We interpret
this as an indication that these soft modes act as strong scatterers for
the magnetophonons and propose that the magnetophonon localization
length is of the order of the average distance between these scatterers,
$R_M \sim [\epsilon_s \nu(\epsilon_s)]^{-1/2}$ (see
Fig.~\ref{Fig_mode}). Using the 3/2-law for the soft mode density of
states $\nu(\epsilon)$, we obtain $R_M \sim (\lambda / \mu)^{5/4} R_c$.
Here and in previous formulas $\lambda = \lambda(1 / R_c)$.

Let us now evaluate $\Delta\omega_p$. To the left-hand side of
Eq.~(\ref{u_L_equation_I}) each scatterer contributes a term of the
order of $(\lambda / R_c^2) (R_c / R_M)^2$, where the first factor is
the characteristic value of $\lambda \nabla^2$ in the vicinity of the
scatterer and the second term is the ratio of the area $\sim R_c^2$
where the nonuniformity in $u_L$ is concentrated to the total
localization area $R_M^2$. The net effect of this term is to shift
$\epsilon$ by a random amount of the order of $\Delta\epsilon \sim
(\lambda / \epsilon_c) (R_c / R_M)^2$. Such shifts produce an
inhomogeneously broadened pinning mode with the quality factor $Q =
\omega_p / \Delta\omega_p \sim \epsilon / \Delta\epsilon \sim (\lambda /
\mu)^{3/2} \gg 1$. This result was previously obtained by a somewhat different
method in Ref.~\cite{Fogler_02} where the lineshape was also calculated:
\begin{equation}
\sigma(\omega) = -i
\frac{e^2 n_e \omega}{m_e \omega_{0}^2}
\frac{1 - i (\omega/ \Omega)^{3/2}}
    {[1 - i (\omega/ \Omega)^{3/2}]^2 - (\omega/\omega_{p})^2}.
\end{equation}
Here $\Omega = \omega_p \sqrt{{\lambda}/{\mu}}$. It should be mentioned
that other theories~\cite{Chitra_98,Fertig_99} give different
predictions for $\Delta\omega_p$. Their critique
is given in Ref.~\cite{Fogler_00}.

\section{Comparison with experiments and conclusions}
\label{Sec:Comparison}

The main uncertainty in comparing the outlined theory with the
experiment is the nature of disorder. It is far from obvious, e.g., why
pinning centers should be weak rather than strong~\cite{Ruzin_92} in
GaAs-based 2DEG. However, assuming this is the case, we can take
empirical values of $\omega_p$ and $n_e$ and work back to the disorder
parameters to see if they are reasonable. In this manner, one obtains
$R_c \sim 6 a$ in samples with largest $Q$~\cite{Fogler_00,Ye_02}. For
the root-mean square amplitude of the random potential we get $U \sim
0.2\,{\rm K}$. As remarked in Sec.~\ref{Sec:Introduction}, the pinning
frequency can also give information about the correlation length $\xi$ of the
random potential. It is easy to
show~\cite{Chitra_98,Fertig_99,Fogler_00} that $\omega_p$ as a function
of $B$ should reach a maximum at the point where $\xi$ is of the order
of the size of the bubbles (a few magnetic lengths). Using this
approach, $\xi$ of the order of a few ${\rm nm}$ can be estimated.
Fertig~\cite{Fertig_99} suggested that the disorder with such
characteristics may originate from surface roughness.

Regarding the linewidth, our theory predicts quality factors $Q \sim
200$ in the limit of infinite sample size and zero temperature. Finite
width of the transmission line ($\sim 30\,\mu{\rm m}$ or $10^3 a$) in
the experiments imposes the upper limit of about 30 on achievable $Q$.
The remaining discrepancy may be related to quantum and thermal effects,
which warrant further study.


\section*{Acknowledgements}

I thank David Huse for previous collaboration on the topics
reported here and Yong Chen, Lloyd Engel, Rupert Lewis, and Dan Tsui
for valuable discussions of the experiments. This work is supported
by Hellman Scholarship Award at University California San Diego.


\begin{thebibliography}{99}

%
%

\bibitem{Das_Sarma_book} M.~Shayegan, Ch. 9
in Perspectives in Quantum Hall Effect,
edited by S.~Das~Sarma and A.~Pinczuk, Wiley, New York, 1997;
H.~A.~Fertig, {\it ibid.\/}, Ch. 3.

\bibitem{Eisenstein_01} For review, see J.~P.~Eisenstein, M.~P.~Lilly,
K.~B.~Cooper, L.~N.~Pfeiffer, and K.~W.~West,
Physica\ E\ {\bf 9} (2001) 1;
J.~P.~Eisenstein,
Solid\ State\ Commun.\ {\bf 117} (2001) 132.

\bibitem{Fogler_review} For review, see M.~M.~Fogler, pp. 98-138 in
High Magnetic Fields: Applications in Condensed
Matter Physics and Spectroscopy, Springer-Verlag, Berlin, 2002;
cond-mat/0111001. For original work, see
A.~A.~Koulakov, M.~M.~Fogler, and B.~I.~Shklovskii,
\prl {\bf 76} (1996) 499;
M.~M.~Fogler, A.~A.~Koulakov, and B.~I.~Shklovskii,
\prb {\bf 54} (1996) 1853; M.~M.~Fogler and A.~A.~Koulakov,
\prb {\bf 55} (1997) 9326.

\bibitem{Li_97} C.-C.~Li, L.~W.~Engel, D.~Shahar, D.~C.~Tsui, and
M.~Shayegan,
\prl {\bf 79} (1997) 1353.

\bibitem{Beya} P.~F.~Hennigan, A.~Beya, C.~J.~Mellor, R.~Gaal,
F.~I.~B.~Williams, and M.~Henini,
Physica\ B\ {\bf 249}-{\bf 251} (1998) 53; A.~S.~Beya,
Ph.D. thesis, L'Universit\'e Paris VI (1998).

\bibitem{Li_00} C.-C.~Li, J.~Yoon, L.~W.~Engel, D.~Shahar,
D.~C.~Tsui, and M.~Shayegan,
\prb {\bf 61} (2000) 10905.

\bibitem{Ye_02} P.~D.~Ye, L.~W.~Engel, D.~C.~Tsui, R.~M.~Lewis, L.~N.~Pfeiffer,
and K.~W.~West,
\prl {\bf 89} (2002) 176802.

\bibitem{Lewis_02} R.~M.~Lewis, P.~D.~Ye, L.~W.~Engel, D.~C.~Tsui,
L.~N.~Pfeiffer, and K.~W.~West,
\prl {\bf 89} (2002) 136804.

\bibitem{Fogler_00} M.~M.~Fogler and D.~A.~Huse,
\prb {\bf 62} (2000) 7553.

\bibitem{Bonsall_77} L.~Bonsall and A.~A.~Maradudin,
\prb {\bf 15} (1977) 1959.
See also R.~C\^ot\'e and
A.~H.~MacDonald,
\prb {\bf 44} (1991) 8759.

\bibitem{Fertig_xxx} R.~C\^ot\'e, C.~Doiron, J.~Bourassa, and H.~A.~Fertig,
cond-mat/0304412.

\bibitem{Gruner_88} G. Gr\"uner,
\rmp {\bf 60} (1988) 1129.

\bibitem{Blatter_94} G.~Blatter, M.~V.~Feigel'man, V.~B.~Geshkenbein,
A.~I.~Larkin, and V.~M.~Vinokur,
\rmp {\bf 66} (1994) 1125.

\bibitem{CDW_pinning_mode} P.~Br\"uesch, S.~Str\"assler, and H.~R.~Zeller,
\prb {\bf 12} (1975) 219;
L.~Degiorgi and G.~Gr\"uner,
\prb {\bf 44} (1991) 7820.

\bibitem{Palaanen_92} M.~A.~Palaanen, R.~L.~Willett, P.~B.~Littlewood,
R.~R.~Ruel, K.~W.~West, L.~N.~Pfeiffer, and D.~J.~Bishop,
\prb {\bf 45} (1992) 11342.



\bibitem{Fukuyama_78} H.~Fukuyama and P.~A.~Lee,
\prb {\bf 18} (1978) 6245.

\bibitem{Normand_92} B.~G.~A.~Normand, P.~B.~Littlewood, and A.~J.~Millis,
\prb {\bf 46} (1992) 3920.

\bibitem{Chitra_98} R.~Chitra, T.~Giamarchi, and P.~Le~Doussal,
\prl {\bf 80} (1998) 3827.

\bibitem{Fertig_99} H.~A.~Fertig,
\prb {\bf 59} (1999) 2120.

\bibitem{Feigelman} V.~M.~Vinokur, M.~B.~Mineev,
and M.~V.~Feigel'man,
Zh.\ Eksp.\ Teor.\ Fiz. {\bf 81} (1980) 2142.
[Sov.\ Phys.\ JETP {\bf 54} (1981) 1138].

\bibitem{Giamarchi_88} T.~Giamarchi and H.~J.~Schulz,
\prb {\bf 37} (1988) 325 and references therein.

\bibitem{Fogler_02} M.~M.~Fogler,
\prl {\bf 88} (2002) 186402.

\bibitem{Gurarie_02} V.~Gurarie and J.~T.~Chalker,
\prl {\bf 89} (2002) 136801.

\bibitem{Aleiner_93} I.~L.~Aleiner and I.~M.~Ruzin,
\prl {\bf 72} (1993) 1056.

\bibitem{Ruzin_92} I.~M.~Ruzin, S.~Marianer, and B.~I.~Shklovskii,
\prb {\bf 46} (1992) 3999.

\end{thebibliography}
\end{document}